\documentclass[preprintnumbers,showpacs,amsmath,amssymb,floatfix,prd,onecolumn,superscriptaddress,nofootinbib]{revtex4}
\usepackage{amssymb}
\usepackage{bbm}
\usepackage{graphicx}
\usepackage{epsfig}
\usepackage{bm}
\usepackage{amsfonts}
\usepackage{epstopdf}
\usepackage{multirow}

\begin{document}

\title{Thermodynamic of the  Ghost Dark Energy Universe}

\author{Chao-Jun Feng}
\email{fengcj@shnu.edu.cn} \affiliation{Shanghai United Center for Astrophysics (SUCA), \\ Shanghai Normal University,
    100 Guilin Road, Shanghai 200234, P.R.China}

\author{Xin-Zhou Li}
\email{kychz@shnu.edu.cn} \affiliation{Shanghai United Center for Astrophysics (SUCA),  \\ Shanghai Normal University,
    100 Guilin Road, Shanghai 200234, P.R.China}

\author{Xian-Yong Shen}
\email{1000304237@smail.shnu.edu.cn} \affiliation{Shanghai United Center for Astrophysics (SUCA),  \\ Shanghai Normal University,
    100 Guilin Road, Shanghai 200234, P.R.China}

\begin{abstract}
Recently, the vacuum energy of the QCD ghost in a time-dependent background is proposed as a kind of dark energy candidate  to explain the acceleration of the Universe. In this model, the energy density of the dark energy   is proportional to the Hubble parameter $H$, which is  the Hawking temperature on the Hubble horizon of the Friedmann-Robertson-Walker (FRW) Universe. In this paper, we generalized this model and choice the Hawking temperature on the so-called trapping horizon,  which will coincides with the Hubble temperature in the context of flat FRW Universe dominated by the dark energy component. We  study the thermodynamics of Universe with this kind of dark energy  and  find that the entropy-area relation is modified, namely, there is an another new term besides the area term.
\end{abstract}

\maketitle
\flushbottom
%###########################################################################%
%Introduction
%###########################################################################%

\section{Introduction}\label{sec:intro}

Current observations converge on the fact that the Universe is accelerated expanding,  lots of theoretical  models were proposed to explain this phenomenology. In some of them, people proposed a new kind of dark energy component with negative pressure  in our Universe, which will drive the acceleration, and the simplest dark energy model is the cosmological constant, but it suffers fine-tuning and coincidence problems. While, in other models, people try to modify the Einstein gravity at large scale in the Universe, e.g. $f(R)$, DGP, etc. models, then the Universe can be accelerated without introducing dark energy.

Recently, a very interesting dark energy model called Veneziano ghost dark energy has been proposed \cite{Urban:2009vy,Urban:2009wb,Urban:2009ke,Urban:2009yg}, and in this model, one can obtain a cosmological constant of just the right magnitude to give the observed expansion from the contribution of the ghost fields, which are supposed to be present in the low-energy effective theory of QCD without introducing any new degrees of freedom.  The ghosts are needed to solve the $U(1)$ problem, but they are completely decoupled from the physical sector \cite{Kawarabayashi:1980dp}. The ghosts make no contribution in the flat Minkowski space, but make a small energy density contribution to the vacuum energy due to the off-set of the cancellation of their contribution in curved space or time-dependent background. For, example, in the Rindler space, the contribution of high frequency modes is suppressed by the factor $e^{-2\pi k/a_T}$ and the main contribution comes from $k\sim a_T$, where $a_T$ is the temperature on the horizon seen by the Rindler observer \cite{Ohta,Cai:2010uf}. In the cosmological context,  one can choice $a_T\sim H$, which corresponds to the temperature on the Hubble horizon. Then, in the context of strongly interacting confining QCD with topological nontrivial sector, this effect occurs only in the time direction and their wave function in other space directions is expected to have the size of QCD energy scale. Thus, this ghost gives the vacuum energy density proportional to $\Lambda_{QCD}^{3} H_{0}$. With  $H_{0}\sim 10^{-33}$eV, it gives the right order of observed magnitude $\sim (3\times10^{-3}$eV)$^{4}$ of the energy density.

It should be emphasized that the Veneziano ghost from the ghost dark energy model is not a new propagating physical degree of freedom and the description of dark energy in terms of the Veneziano ghost is just a matter of convenience to describe very complicated infrared dynamics of strongly coupled QCD and it does not violate unitarity, causality, gauge invariance and other important features of renormalizable quantum field theory, see \cite{Zhitnitsky:2010ji,Holdom:2010ak,Zhitnitsky:2011tr,Zhitnitsky:2011aa}. Generally, it is very difficult to accept the linear behavior that the energy of FRW Universe is linear in Hubble constant "$H$", because QCD is a theory with a mass gap determined by the energy scale $100$MeV. So, it is generally expected that there should be an exponentially small corrections rather than that linear corrections $H$. However, as the arguments discussed in refs.\cite{Zhitnitsky:2010ji, Holdom:2010ak,Zhitnitsky:2011tr,Zhitnitsky:2011aa} that the linear scaling $H$ is due to the complicated topological structure of strongly coupled QCD, not related to the physical massive propagating degree of freedom. Therefore, the linear in Hubble constant $H$ scaling has a strong theoretical support tested in a number of models. The recent progress on the GDE model, see Ref.~\cite{Cai:2010uf2, Feng:2012wx, Ebrahimi:2011js}, in which the author also consider other possibilities of the energy density form of the GDE model with or without interactions and also its thermodynamical behaviors.

However, there are some other horizons in the context of the FRW Universe, such as the future inner trapping horizon, which will be defined in the next section. One can also use the outer trapping horizon  to  define  black holes in a general spacetime including time-dependent spacetime.  The Hawking radiation of apparent horizon in a FRW Universe was firstly studied in Ref.\cite{Cai:2008gw}. In this paper, we will choice $a_T\sim T_{t}$, where $T_{t}$ is the temperature defined on the trapping horizon of the  FRW Universe. We study its thermodynamic behavior, and find that there is a new term contributes to the entropy-area relation.

On the other side, there exits some kind of correspondence between thermodynamical laws and gravitational equations. We will give a briefly review on this topic in the following section. In this paper, we assume that the thermodynamic laws as the origin of all the dark energy and modification gravity models, then we do the comparison between the ghost dark energy model (including the generalized ones) and DGP model, and find  they are very similar but still have differences from the thermodynamical point of view.

This paper is organized as follows: In Sec.~\ref{sec:firstlaw}, we give a briefly review on the unified first law of thermodynamics on the ``inner'' trapping horizon. In Sec.~\ref{sec:DE}, we study the thermodynamics of the ghost dark energy Universe and some generalized cases on the trapping horizon. Also we get the property of the generalized second law of thermodynamics of this model in this section.   In the last section, we will give some discussions and conclusions.

%###########################################################################%
%
%A brief review of the First Law
%
%###########################################################################%

\section{Briefly review on the unified first law}\label{sec:firstlaw}

In this section, we will give a  brief review on the "unified first law" in the $3+1$-dimensional spherical symmetric spacetime.  This "unified law" was first proposed by Hayward \cite{Hayward:1997jp,Hayward:1993wb,Hayward:1994bu,Hayward:1998ee}, and it is a thermodynamical description of Einstein equations. It indicates that there could be some relations between thermodynamic laws and gravitational equations. And it seems that the gravitational theory maybe not a fundamental theory. Then, one can just use the thermodynamic laws to describe the gravity.  For convenient, we will use the double-null form of the metric, namely, $ds^{2}=-2e^{-f}d\xi^{+}d\xi^{-}+r^{2}d\Omega^{2}$. Here, $d\Omega^{2}$ is the line element of the $2$-sphere with unit radius, $r$ and $f$ are functions of ($\xi^{+},\xi^{-}$). Each symmetric sphere has two preferred normal directions, namely the null directions $\partial /\partial\xi^{\pm}$, which will be assumed future-pointing in the following.  And also, we will assume the spacetime is time-orientable.

The expansions of the radial null geodesic congruence are defined by $\theta_{\pm}=2r^{-1}\partial_{\pm}r$, and $\partial_{\pm}$ denotes the coordinates derivative along $\xi^{\pm}$. The expansion measures whether the light rays normal to the sphere are diverging $(\theta_{\pm} > 0)$ or converging $(\theta_{\pm} < 0)$, namely, whether the sphere is increasing or decreasing in the null directions. It should be noticed that the sign of $\theta_{\pm}$ will not change with geometries, while its value will. The only invariants of the metric and its first derivative are functions of $r$ and $e^f \theta_{+}\theta_{-}$, or equivalently $g^{ab}\partial_{a}r\partial_{b}r = -\frac{1}{2}e^{f}\theta_{+}\theta_{-}$, which has an important physical and geometrical meaning: a sphere is said to be trapped (untrapped), if $\theta_{+}\theta_{-} >0  ( \theta_{+}\theta_{-} <0 ) $.  And it is called a marginal sphere if $\theta_{+}\theta_{-}=0$.

In the case of non-stationary black holes, Hayward \cite{Hayward:1997jp} has proposed that the future outer trapping horizons defined as the closure of a hypersurface foliated by future or past, outer or inner marginal sphere is taken as the definition of black holes, since the horizon possess various properties which are often intuitively ascribed to black hole including confinement of observers and analogues of the zeroth, first and second law of thermodynamics. However, in the case of FRW Universe,  one should take the  future inner trapping horizon defined by \cite{Cai:2005ra,Cai:2006rs,Akbar:2006mq,Akbar:2006er,Cai:2008ys,Cai:2008mh, Feng:2011xi,Feng:2011li}, $\theta_{+} = 0\,, \quad \theta_{-} < 0\,, \quad \partial_{-}\theta_{+} >0 $
 as a system on which the thermodynamics will be established, since the surface gravity is negative on the cosmological horizon.

The $(0,0)$ component of Einstein equations could be written as a ``unified first law'' on this future inner trapping horizon as \cite{Cai:2005ra}
\begin{equation}\label{uflofbh}
    \langle dE,z \rangle =  \frac{\kappa}{8\pi G}\langle dA,z \rangle + \langle  WdV,z \rangle   \,,
\end{equation}
where $E$ is defined as the Minsner-Sharp energy, $A = 4\pi r^{2}$ and $V = \frac{4\pi}{3} r^{3}$. Here,  we have defined  two invariants constructed from the  energy-momentum tensor $T^{\mu\nu}$:
\begin{eqnarray}
  W &=&  - \frac{1}{2} g_{ab}T^{ab} \,   = -g_{+-}T^{+-} \,, \label{wd}\\
  \Psi_{a} &=& T^{b}_{a} \partial_{b} r + W \partial_{a} r , \label{efv}
\end{eqnarray}
which are called the work density, and   the energy flux vector (also called the energy-supply  vector).
Here and in the following,  $a, b$ denotes the two dimension space normal to the sphere. Note that, the Minsner-Sharp energy is a pure geometric quantity and has much better properties than the other definitions of energy  when one consider the case of non-stationary spacetime. The relation between the  Minsner-Sharp energy and others could be found in Ref.~\cite{Hayward:1997jp} .

In Eq.~(\ref{uflofbh}), $\kappa$ is defined by
\begin{equation}\label{surg}
    \kappa = \frac{1}{2} \nabla^{a} \nabla_{a} r \,,
\end{equation}
 which is called the surface gravity of the trapping horizon, and $z$ is a vector tangent to the trapping horizon \cite{Feng:2011xi,Feng:2011li}, which satisfies $z^{a}\partial_{a}(\partial_{+}r) = 0$ on the trapping horizon. By taking the Einstein equations  $\partial_{+}\partial_{+} r = -4\pi r T_{++}$, see Ref.~\cite{Hayward:1997jp} and the definition of the surface gravity in Eq.~(\ref{surg}), one can easily finds
\begin{equation}\label{eq:heat_flux}
     \langle A\Psi,z \rangle = \frac{\kappa}{8\pi G}\langle dA,z \rangle \,,
\end{equation}
which is the Clausius relation in the version of black hole thermodynamics, see the first term on the right side of  Eq.~(\ref{uflofbh}). The left side of the above equation is nothing but the heat flow $\delta Q$, and the right side has the form $TdS$, if one identities the temperature $T=|\kappa|/2\pi$ and the  entropy $S = A/4G$. So, in Einstein theory, the ``unified first law'' also implies the Clausius relation \cite{Cai:2005ra,Cai:2006rs,Akbar:2006mq,Akbar:2006er,Cai:2008ys,Cai:2008mh, Feng:2011xi,Feng:2011li}.

%###########################################################################%
%
%Thermodynamics of the ghost dark energy Model
%###########################################################################%

\section{Thermodynamics of the ghost Dark Energy Model}\label{sec:DE}

In FRW Universe, one can also write the metric in the double-null form
\begin{equation}
    ds^2=-2 d\xi^+d\xi^- +\tilde{r}^2d\Omega^2 \,,
\end{equation}
where $\tilde{r}=a(t)r$, and $a(t)$ is the scale factor. The definition of $\xi^{\pm}$ are the same as that in \cite{Cai:2005ra,Cai:2006rs,Akbar:2006mq,Akbar:2006er,Cai:2008ys,Cai:2008mh, Feng:2011xi,Feng:2011li}. By solving the equation $\partial_{+} \tilde r |_{\tilde r= \tilde r_{T}} = 0$, we obtain the trapping horizon $\tilde r_{T}$  as
\begin{equation}\label{horizon_radus}
    \tilde r_{T} = \left(H^{2}+ \frac{k}{a^{2} } \right )^{-1/2} \,,
\end{equation}
which coincides with the apparent horizon. The surface gravity is now $\kappa = -(1-\epsilon)/\tilde r_{T}$, where  we have defined $\epsilon \equiv \frac{\dot{\tilde r}_{T}}{2H \tilde r_{T}}$. It is easy to check that $\partial_{-}\tilde r_{T} <0$, and the trapping horizon is future. Then, the project vector is given by $z= \partial_{t} - (1-2\epsilon)Hr \partial_{r}$, when $z^{+} = 1$ is chosen in the $(t, r)$ coordinates.

Actually, when one applies the first law to the apparent horizon to calculate the surface gravity and thereby the temperature by considering an infinitesimal amount of energy crossing the apparent horizon, the apparent horizon radius $\tilde r_{T}$ should be regarded to have a fixed value \cite{Cai:2005ra}, so in this sense, one has $\kappa \approx -1/\tilde r_{T} $. And also, one can see in the following, $\epsilon$ could be neglected when dark energy dominates the Universe. Therefore, the energy density of the ghost dark energy is given by as follows
\begin{equation}\label{omodel}
    \rho_{DE} = \alpha \sqrt{H^{2} + \frac{k}{a^{2}}} \,,
\end{equation}
 where $\alpha>0$ and is roughly of order of $\Lambda_{QCD}^3$ as we mentioned in the introduction section. Here, all the proportional coefficients are also absorbed in $\alpha$, whose value could be given by observations, see the work of Cai et al. in Ref.~\cite{Cai:2010uf}.

 The Friedmann equation reads
 \begin{equation}\label{eq:friedeq1}
          H^{2}+\frac{k}{a^{2}}=\frac{1}{3}\left( \rho_{DE}+\rho_{m}\right)\,,
 \end{equation}
 where we have set $8\pi G=1$.
By solving the above equation and the continuity equations, we obtain the energy density and pressure of the dark energy
\begin{eqnarray}
  \rho_{DE} &=&  \frac{\alpha^{2}}{6} \left(1+ \sqrt{1+\frac{12\rho_{m}}{\alpha^{2}}}\right) \,, \\
  p_{DE} &=& \rho_{m}\left( 1+\frac{12\rho_{m}}{\alpha^{2}}\right)^{-1/2}-\rho_{DE} \,, \label{eq:pDE}
\end{eqnarray}
 where we have neglected the $p_{m}\approx 0$ for matter.  The equation of state parameter of the dark energy is
 \begin{equation}\label{eosde}
    w_{DE} = \frac{p_{DE}}{\rho_{DE}} = -\frac{1}{2}\left[1+\left(1+\frac{12\rho_{m}}{\alpha^{2}}\right)^{-1/2}\right] \,.
\end{equation}
When the matter decays $\rho_{m}\sim a^{-3}=(1+z)^{3}$, the dark energy will dominate the Universe, and its equation of state will trend to $w_{DE}\rightarrow -1$, see Eq.~(\ref{eosde}). And the present value of the equation of state is
\begin{equation}
    w_{DE0} = -1+ \frac{\Omega_{m0}}{\Omega_{DE0}+2\Omega_{m0}}
    + \frac{3\Omega_{m0}\Omega_{DE0}(\Omega_{m0}+\Omega_{DE0})}{(\Omega_{DE0}+2\Omega_{m0})^{3}} z + o(z)\,.
\end{equation}
From the Friedmann equation (\ref{eq:friedeq1}), we also have
 \begin{equation}
    \alpha = 3H_{0}\Omega_{DE0} (\Omega_{DE0}+\Omega_{m0})^{-1/2} M_{pl}^{2} \,,
\end{equation}
where $\Omega_{m0}=\rho_{m0}/(3H_{0}^{2})$, $\Omega_{DE0} = \rho_{DE0}/(3H_{0}^{2})$. Taking the values of $\Omega_{m0}\approx0.27$ and $\Omega_{DE0}\approx0.73$, one can get $w_{DE0}\approx -0.813+0.286z$ and $\alpha \approx(100$ Mev$)^{3}$. This values are consistent with the recent observations: $w^{obs}_{DE0} = -0.93^{+0.13}_{-0.13} + \left(0.41^{+0.71} _{-0.72} \right)z \,, $ ($68 \% $CL) in curved Universe ($k\neq0$) from WMAP+BAO+H$_{0}$+SN \cite{Komatsu:2010fb}.

By using the definition of $\epsilon$ and the Friedmann equation without approximation of $\kappa \sim 1/\tilde r_{T}$, we obtain the following relation
\begin{equation}\label{eosr}
    \epsilon = \frac{3}{4}\left(1+ \frac{w_{DE}}{1+ r_{c}}\right) \,, \quad r_{c}= \frac{\rho_{m}}{\rho_{DE}} \,.
\end{equation}
For the detailed calculations, see Appendix \ref{eps}. From Eq.~(\ref{eosr}), it shows that  when the dark energy dominates the Universe ($r_{c}\rightarrow 0$, $w_{DE}\rightarrow -1$),  $\epsilon $ can be neglected.  Then, in this sense our approximation for the energy density of the ghost dark energy (\ref{omodel}) is also  reasonable.

From Eqs.~(\ref{wd}) and (\ref{efv}), we obtain the  work density $W_{e}$ and $\Psi_{e}$ for the dark energy as
\begin{eqnarray}
    W_{e} &=& \rho_{DE} - \frac{\rho_{m}}{2}\left(1+\frac{12\rho_{m}}{\alpha^{2}}\right)^{-1/2} \,, \\
    \Psi_{e} &=&  \frac{\rho_{m}}{2}\left(1+\frac{12\rho_{m}}{\alpha^{2}}\right)^{-1/2} \bigg(-H\tilde r_{T} dt + a dr \bigg)  \,,
\end{eqnarray}
and then we have
\begin{equation}\label{heatf2}
    \delta Q_{DE}   =  \langle A\Psi_{DE}, z\rangle
    =  \frac{2\kappa A H \alpha^{2}\epsilon }{3\rho_{DE}} =\frac{\pi \alpha^{2}}{3\rho_{DE}} T\langle dA, z\rangle\,,
\end{equation}
where we have used the relation (\ref{f31}) and (\ref{f32}). Here, $A$ denotes the surface area of the trapping horizon, namely, $A=4\pi \tilde r_{T}^{2}$. For the  heat flow of the pure matter $\rho_{m}$, we also have
\begin{equation}
      \delta Q_{m} = \frac{\kappa}{8\pi G}\langle dA, z\rangle - \langle A\Psi_{DE}, z\rangle
      = 2\pi \left(1- \frac{\alpha}{12\sqrt{\pi}} A^{1/2}\right)T\langle dA, z\rangle = T\langle dS_{m}, z\rangle  \,,
\end{equation}
where we have used the relation $A = 4\pi \alpha^{2}/\rho_{DE}^{2}$ from the Friedmann equation.
Therefore, the entropy is given by
\begin{equation}\label{entro ghost}
    S_{m} = 2\pi A - \frac{\sqrt{\pi} \alpha}{9} A^{3/2} \,,
\end{equation}
up to some integration constant. Here, the first term on the right hand side of the above equation is nothing but $A/4G$ when one recovers the induced Planck mass, while the second term is the additional term that becomes important when  $A\gtrsim M_{pl}^{4}\alpha^{-2} \sim H_{0}^{-2}$.

As we known, in the Dvali-Gabadadze-Porrati (DGP) model, the entropy is given by  \cite{Wang:2009,Wang:2007}
\begin{equation}\label{entro 31}
    S_{m}=\frac{A}{4G}\pm\frac{1}{24\sqrt{\pi} r_{c}G} A^{\frac{3}{2}}\,,
\end{equation}
 where $r_{c} = G_{5}/(2G)$ is the cross-over scale  in the DGP model and $G_{5}$ is the $5$-dimension gravitational constant in the bulk. Here, the minus sign in Eq.~(\ref{entro 31}) corresponds to the self-accelerating branch of the DGP model, while plus sign corresponds to normal ( non-accelerating ) branch. Therefore, the entropies of the ghost dark energy model and DGP model (self-accelerating branch) are of the same order if $\alpha \sim 1/G_{5}$.

 \section{Generalized case}

 It is also interesting to consider higher order terms in the energy density of the dark energy, namely, Eq.~(\ref{omodel}) can be generalized to
 \begin{equation}\label{hot}
    \rho_{DE}=\alpha\sqrt{H^{2}+k/a^{2}}+ \beta(H^{2}+k/a^{2})  \,,
\end{equation}
and when $\beta \rightarrow 0$, we recovered the model discussed in Sec.~\ref{sec:DE}. Eq.~(\ref{hot}) was first proposed in Ref. \cite{Grande:2009,Grande:20091,Grande:20092} to get an accelerating Universe. By solving the Friedmann equation, we obtain
\begin{equation}
    \rho_{DE} = \frac{3\alpha^{2}}{2(3-\beta)^{2}} \left(1\pm \sqrt{1+\frac{4(3-\beta)\rho_{m}}{\alpha^{2}}}\right) + \frac{\beta}{3-\beta} \rho_{m} \,,
\end{equation}
for $\beta \neq 3$, while  $\rho_{DE} = 3\alpha^{-2} \rho_{m}^{2} -\rho_{m}$
for $\beta = 3$ (and it requires $\alpha<0$). Then, in the case of $\beta\neq3$, the associated work density $W_{e}$ and $\Psi_{e}$ for the dark energy as
\begin{eqnarray}
    W_{e} &=& \rho_{DE} - \frac{\rho_{m}}{2(3-\beta)} \left[ \beta \pm 3\left( 1+\frac{4(3-\beta)\rho_{m}}{\alpha^{2}}\right)^{-1/2}\right]\,, \\
    \Psi_{e} &=& \frac{\rho_{m}}{2(3-\beta)} \left[ \beta \pm 3\left( 1+\frac{4(3-\beta)\rho_{m}}{\alpha^{2}}\right)^{-1/2}\right]
    \bigg(-H\tilde r_{T} dt + a dr \bigg)  \,,
\end{eqnarray}
and then we get
\begin{equation}\label{heatf3}
    \delta Q_{DE}   =\frac{\pi }{3} \left( \frac{\alpha A^{1/2}}{2\sqrt{\pi}} + 2\beta\right) T\langle dA, z\rangle\,,
\end{equation}
and
\begin{equation}
      \delta Q_{m} = 2\pi \left(1-\frac{\beta}{3}- \frac{\alpha}{12\sqrt{\pi}} A^{1/2}\right)T\langle dA, z\rangle   \,,
\end{equation}
for the  heat flow of the pure matter $\rho_{m}$. Therefore, the entropy is given by
\begin{equation}\label{entro 38}
    S_{m} = 2\pi \left(1-\frac{\beta}{3}\right) A - \frac{\sqrt{\pi} \alpha}{9} A^{3/2} \,,
\end{equation}
up to some integration constant. From Eq.~(\ref{entro 38}), one can see that the effect of $\beta$ is to modify the slope of the entropy area relation in the first term of the right hand side of the above equation, while it does not contribute to the second term.  In the case of $\beta = 3$, we have
\begin{equation}
    W_{e} = -\frac{\rho_{m}}{2}\,, \qquad
    \Psi_{e} = \left(3\alpha^{-2}\rho_{m}^{2} - \frac{\rho_{m}}{2} \right)
    \bigg(-H\tilde r_{T} dt + a dr \bigg)  \,,
\end{equation}
and then
\begin{equation}\label{heatf4}
    \delta Q_{DE}   =2\pi \left( 1- \frac{\alpha^{2}}{6\rho_{m}}\right) T\langle dA, z\rangle\,, \quad
    \delta Q_{m} =  -\frac{\alpha \sqrt{\pi}}{6} A^{1/2} T\langle dA, z\rangle \,.
\end{equation}
Therefore, the entropy is given by
\begin{equation}
    S_{m} =  - \frac{\sqrt{\pi} \alpha}{9} A^{3/2} \,,
\end{equation}
up to some integration constant. Here, $\alpha <0$, so the entropy increase with area.

We can also generalize the ghost dark energy model in a more general case with the energy density
\begin{equation}\label{gen case}
    \rho_{DE} = \sum_{n=-l}^{m} \alpha_{n} \left(H^{2}+\frac{k}{a^{2}}\right)^{n/2}  \,, \quad m,l \ge 0 \,.
\end{equation}
Although  we can not obtain the exact relation between $\rho_{DE}$ and $\rho_{m}$, we can still get the entropy area relation by using the same approach
\begin{equation}\label{entro gen ok}
    S_{m} = 2\pi \left(1-\frac{\alpha_{2}}{3}\right) A - \frac{16\pi^{2}}{3} \alpha_{4}\ln A
    - \frac{2\pi}{3} \sum_{\substack{n=-l, \\n \neq 0,2,4}} ^{m} \frac{n\alpha_{n}}{4-n}  (4\pi)^{\frac{n-2}{2}} A^{\frac{4-n}{2}} \,,
\end{equation}
up to some integration constant. The logarithm term of the above equation could be  also obtained  from loop quantum cosmology \cite{Cai:2005ra}, and usually, this term is regarded as a quantum correction. It should be noticed that the constant term $\alpha_{0}$ (cosmological constant) do not contribute to the above entropy area relation.

Before ending this section, we would like to say something about the second law of thermodynamics of this model. The time derivative of entropy is given by
\begin{equation}
    \dot S_{m} = 8\pi^{2} H \tilde r_{T}^{4} (\rho_{m} + p_{m})  \,.
\end{equation}
In our cases, we have neglected the pressure of matter $p_{m}\approx 0$, so we have $\dot S_{m}\geq 0$ which guarantees the second law of thermodynamics for our models.  Furthermore, when dark energy dominates the Universe ($\rho_{m} \approx 0$), the entropy could be almost a constant depending on the parameters $\alpha_{n}$.

\section{Discussion and Conclusion}
In this paper, we have studied the thermodynamics of the Universe with the   Veneziano ghost dark energy component to drive the acceleration.  By using the unified first law, we obtain the temperature and the corresponding entropy on the trapping horizon, which coincides with the apparent horizon in the FRW Universe. We find that there is a new term contributes to the entropy-area relation, see Eq.~(\ref{entro ghost}). This relation is very similar to that from the DGP model in the self-accelerating branch, so it can not distinguish these two models at this background evolution level, but it may be to distinguish them by doing the perturbation theory, and we will do the further studies on this subject.

We also generalized this model by adding a higher order term in Eq.~(\ref{hot}). We solve the Friedmann equation exactly, and find the corresponding entropy area relation on the trapping horizon. Now, there are two new terms   contribute to the relation. For a more generalized case , we have obtained the exact entropy area relation, which includes a logarithm term that could be also obtained  from loop quantum cosmology and many terms with powers of area. However, there is no quadratic term $\sim A^{2}$, and the cosmological constant do not contribute this relation also.

As we have mentioned, there is a relation between gravity theory and thermodynamics, and gravity may be not a fundamental theory. Therefore, we can regard  the thermodynamical laws as the first principle to get a physical model.  For example, if we start from a general form of entropy area relation like Eq.~(\ref{entro gen ok}), we can get the dark energy model with the same background equations as the generalized  Veneziano ghost dark energy model, since there are could be some other models that have the same entropy area relation. But, one can still considers  the thermodynamics laws as the origin of such kind of models.

Also, we find that the time derivative of entropy can not be negative, which means the second law of thermodynamics is hold in our models. Furthermore, If one assumes that some other components with equation of state $w<-1$ also exit in our models, then a generalized second law of thermodynamics is needed, see Ref.\cite{Wang:2009}, and it deserves further studies.

\acknowledgments

This work is supported by National Science Foundation of China grant Nos.~11105091 and~11047138, National Education Foundation of China grant  No.~2009312711004, Shanghai Natural Science Foundation, China grant No.~10ZR1422000, Key Project of Chinese Ministry of Education grant, No.~211059,  and  Shanghai Special Education Foundation, No.~ssd10004.

\appendix
\section{Calculation $\epsilon$ without approximation}\label{eps}
Without the approximation $\kappa\sim 1/\tilde r_{T}$, the energy density of the dark energy is given by
\begin{equation}
    \rho_{DE} = \frac{\alpha}{\tilde r_{T}} \left(1-\epsilon\right) \,,
\end{equation}
and from the Friedmann equation, we get
\begin{equation}\label{f31}
    \dot H -\frac{k}{a^{2}}= -\frac{1}{2} \bigg[ \rho_{DE}(1+w_{DE}) + \rho_{m}\bigg] \,,
\end{equation}
while we also have the relation
\begin{equation}\label{f32}
    \dot H -\frac{k}{a^{2}} = -\frac{2\epsilon}{\tilde r_{T}^{2}} \,,
\end{equation}
by using the definition of $\epsilon$. From Eqs.~(\ref{f31}) and (\ref{f32}), we get the following relation
\begin{equation}
    \frac{4\epsilon}{3} = 1+\frac{w_{DE} }{1+r_{c}}  \,,
\end{equation}
where $r$ is the ratio of energy density of dark energy to dark matter, and this is just Eq.~(\ref{eosr}).

\end{document}